\documentclass[12pt]{iopart}
\usepackage[dvips]{graphicx}
\usepackage{hyperref}
\hypersetup{ pdftitle={Dielectric multilayer waveguides for TE and
TM mode matching}, pdfauthor={Dzmitry M. Shyroki, Andrei V.
Lavrinenko}, pdfkeywords={Planar waveguides; dispersion relation;
perfect phase matching; mode crossings} }
\begin{document}
Published in: J. Opt. A: Pure Appl. Opt. \textbf{5} (2003)
192--198

Online at: \url{stacks.iop.org/JOptA/5/192}

\title{Dielectric multilayer waveguides for TE and TM mode matching}
\author{D M Shyroki and A V Lavrinenko}
\address{Physics Faculty, Belarusian State University,
        Fr. Scaryna Avenue 4, 220080 Mensk, Belarus}
\ead{shyroki@tut.by}
%%%%%%%%%%%%%%%%%%%%%%%%%%%%%%%%%%%%%%%%%%%%%%%%%%%%%%%%%%%%%
\begin{abstract}
We analyse theoretically for the first time to our knowledge the
perfect phase matching of guided TE and TM modes with a multilayer
waveguide composed of linear isotropic dielectric materials.
Alongside strict investigation into dispersion relations for
multilayer systems, we give an explicit qualitative explanation
for the phenomenon of mode matching on the basis of the standard
one-dimensional homogenization technique, and discuss the minimum
number of layers and the refractive index profile for the proposed
device scheme. Direct applications of the scheme include
polarization-insensitive, intermodal dispersion-free planar
propagation, efficient fibre-to-planar waveguide coupling and,
potentially, mode filtering.

As a self-sufficient result, we present compact analytical
expressions for the mode dispersion in a finite, $N$-period,
three-layer dielectric superlattice.\\
\\
\textbf{Keywords:} Planar waveguides, dispersion relation, perfect
phase matching
\end{abstract}

%\pacs{42.82.Et, %Waveguides, couplers, and arrays (in Integrated optics)
%      42.79.Gn, %Optical waveguides and couplers
%      42.25.Ja, %Polarization
%      42.25.Lc, %Birefringence
%      }

\maketitle
%%%%%%%%%%%%%%%%%%%%%%%%%%%%%%%%%%%%%%%%%%%%%%%%%%%%%%%%%%%%%
\section{Introduction}
Since the early experiments on mode conversion in film-waveguide
magneto-optical systems~\cite{Tien74,Bruns75} and the subsequent
theoretical works~\cite{Gillies76,Hlawiczka78a,Hlawiczka78b} it
has become clear that an equality between propagation constants of
the two guided modes---otherwise called mode matching or perfect
phase matching---can be realized exclusively due to the tensorial
character of dielectric permittivity of the waveguiding slab, i.e.
due to its either natural or induced anisotropy. An obvious way to
gain anisotropy is to produce a waveguide layer of monocrystalline
material possessing intrinsic birefringence; another possibility
is provided by use of gyrotropic waveguides (for example, YIG on
GGG) subject to an external DC magnetic field. Historically, it
was the latter way that experimentalists followed, probably
because of its higher flexibility and technical reliability. A
number of device schemes implementing that principle were rapidly
proposed, from an optical switch~\cite{Tien74} to a sensitive
experimental method of determining the permittivity tensor
elements~\cite{Gillies76}. The two unpleasant drawbacks were then
the complexity of the schemes and the comparatively high
absorption in gyrotropic films that hindered their utilization,
e.g. for intermodal dispersion-free planar propagation.

On the other hand, in recent years a certain interest in
traditional multilayer dielectric mirrors from the viewpoint of
photonic bandgap localization has been observed, stimulated
probably by the unexpected discovery of total omnidirectional
reflection from an appropriately designed one-dimensional
dielectric lattice, i.e. an actual photonic band gap exhibited in
case of such a simple
geometry~\cite{Winn,Dowling,Chigrin,Russell}. What looks now
curious but is important for the purpose of the present treatment
is that one can easily see the points of perfect phase matching of
TE and TM modes on the standard band structure diagrams of
dielectric multilayers \cite[ch~4]{Joannopoulos}, yet
unfortunately little or no attention has been paid in the
literature to those mode matching phenomena---in spite of so many
promising applications including polarization-insensitive,
intermodal dispersion-free planar waveguiding, efficient
fibre-to-planar waveguide coupling and, potentially, mode
filtering, drastic improvement of mode conversion efficiency and
diffraction efficiency in magneto-optical Bragg cells and others.
To the best of our knowledge, the underlying idea that a
multilayer waveguide can be regarded as a homogeneous, effectively
anisotropic slab---and as if it were the case of a naturally
birefringent or gyrotropic, magnetized waveguide, the perfect
phase matching of TE and TM modes must occur---has been formulated
just recently~\cite{me}.

The aim of the current paper is to highlight and theoreticallly
investigate the perfect phase matching of guided TE and TM modes
with a multilayer device composed of linear, isotropic,
non-magnetic and non-absorbing materials. The paper is organized
as follows. \hyperref[WhatIs]{Section~\ref*{WhatIs}} is devoted
mainly to qualitative models and considerations: we begin with a
transparent explanation for the mode matching phenomenon on the
basis of conventional one-dimensional homogenization method and
then discuss the general eigenproblem for TE and TM modes in order
to derive some basic requirements for the scheme, such as the
minimum number of layers and the proper refractive index profile.
Strict analytical expressions for the modal dispersion structure
of periodic dielectric stacks with two-layer and three-layer unit
cells follow those qualitative speculations and are given in
\hyperref[Dispersion]{section~\ref*{Dispersion}}. Finally, we
analyse in \hyperref[Simulated]{section~\ref*{Simulated}} several
examples of perfect phase matching with simple planar structures
composed of silicon and silicon-nitride layers, fully compatible
with modern integrated optics techniques.

%%%%%%%%%%%%%%%%%%%%%%%%%%%%%%%%%%%%%%%%%%%%%%%%%%%%%%%%%%%%%

\section{\label{WhatIs}What is mode matching by a multilayer dielectric waveguide?}
A straightforward way to make the idea of perfect phase matching
transparent is to consider a stack of isotropic dielectric layers
as an effectively homogeneous, uniaxially birefringent medium, by
means of the standard one-dimensional homogenization technique.
This approach allows to expand the argumentation for the mode
matching with anisotropic waveguides to the case of dielectric
multilayers, with an important consequence that the mode matching
is in principle possible with those structures too. To find the
minimum number of layers required for the matching and to optimize
the refractive index profile, we furthermore analyse the
waveguiding conditions for TE and TM modes in multilayer
structures, tracking the fruitful analogy with confined
one-dimensional electron states problem quite profoundly
investigated in the framework of classical quantum mechanics and
solid-state physics.

\subsection{Effective medium treatment}
\begin{figure}
\begin{center}
\includegraphics[scale=0.9]{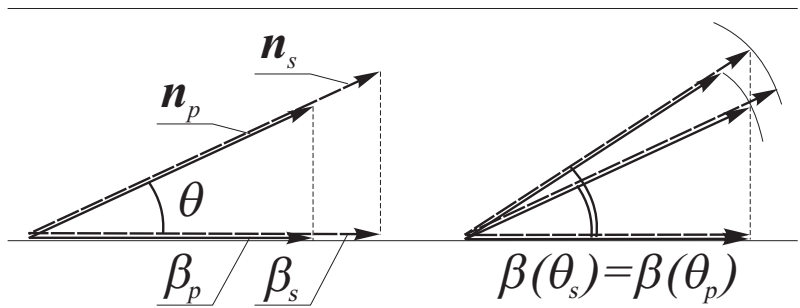}
\end{center}
\caption{Refraction vectors $\bi{n}_{s,p} = n_{s,p}\tilde{\bi{k}}$
($\tilde{\bi{k}}$ is the wave normal) and propagation constants
$\beta_{s,p}$ in a birefringent waveguiding medium.} \label{f1}
\end{figure}

In a single-layer isotropic waveguide, the two given modes---TE
(s-polarized) and TM (p-polarized)---are commonly characterized at
a fixed frequency by definitely different propagation angles
$\theta_{s,p}$ between the wavevector and the plane of the guide,
and consequently, by different (non-equal) propagation constants
$\beta_{s,p}$ via
\begin{equation}
\beta_{s,p} = n\,\cos{\theta_{s,p}}\,,
\end{equation}
where $n$ is the refractive index of the waveguiding medium and
the subscripts $s$ and $p$ differentiate between TE and TM modes.
Anisotropy of the waveguide accounts evidently for the following
modification of equation (1):
\begin{equation}
\beta_{s,p} = n_{s,p}(\theta)\,\cos{\theta_{s,p}}\,,
\end{equation}
where $n_{s,p}$ are the indices `seen' by TE and TM modes, and a
situation like that in \hyperref[f1]{figure~\ref*{f1}} becomes
possible, when the projections of the wave vectors of different
modes onto the plane of guidance coincide, giving rise to the
phenomenon of mode matching.

Consider now a two-component periodic multilayer system consisting
of layers of dielectric permittivity $\epsilon_1=n_1^2$ and
thickness $t_1$ separated by layers with permittivity
$\epsilon_2=n_2^2$ and thickness $t_2$. If a plane wave is
incident onto this array with its electric vector $\bi{E}$
polarized parallel to the layers (TE-wave) and the thicknesses
$t_{1}$ and $t_{2}$ are small compared to the wavelength, then one
may introduce the effective dielectric permittivity of the array
as~\cite{BornWolf}
\begin{equation}
\epsilon_s = \frac{t_1\,\epsilon_1 + t_2\,\epsilon_2}{t_1 + t_2},
\end{equation}
and if the magnetic vector $\bi{H}$ is parallel to the plates (TM-wave), then the effective
permittivity becomes~\cite{BornWolf}
\begin{equation}
\epsilon_p = \frac{(t_1 + t_2)\,\epsilon_1\,\epsilon_2}{t_1\,\epsilon_1 + t_2\,\epsilon_2}.
\end{equation}

This treatment allows one to state that two isonormal, s- and
p-polarized waves `see' in general different refractive indices of
effective waveguiding medium, hence if to start from the isotropic
($n_1 = n_2$) case and then change one of the indices, the
families of TE and TM dispersion curves will undergo non-equal
shifts on the dispersion structure diagram and thus intersect at a
number of points, displaying the modes degeneration---the perfect
phase matching. One should remember by the way that it is of
course not complete, spatial degeneration of modes, but mere
propagation constant coincidence.

The question that naturally arises here is: How much layers are
enough for the mode matching to occur? In other words: When does
the reasoning based on the form birefringence approach lose its
validity? To answer these questions and to elucidate the physics
of mode matching in the case of a finite number of layers, one may
find it convenient to employ an explicit mathematical analogy
between confined photon and electron states, the latter being
quite thoroughly investigated in classical quantum mechanics and
solid-state physics---a comprehensible 1D Kronig--Penney model
should certainly be mentioned here as an example.

\subsection{Dispersion equation for TE and TM modes as an eigenproblem: Qualitative analysis}
As is commonly known and excellently described by Joannopoulos
\etal~\cite[ch~3]{Joannopoulos}, symmetry considerations allow us
to separate the modes of planar dielectric waveguides into two
classes: TE and TM polarized. In a coordinate system in which the
$x$-axis is normal to the bimedium interfaces and the light
propagates along the $z$-direction, we can write for TE and TM
polarizations, respectively,
\begin{equation}
\bi{E}(\bi{r}) = \exp(\rmi \frac{\omega}{c} \beta z) \phi^E(x) \bi{y}
\end{equation}
and
\begin{equation}
\bi{H}(\bi{r}) = \exp(\rmi \frac{\omega}{c} \beta z) \phi^H(x) \bi{y},
\end{equation}
where $\omega$ is the angular frequency, $c$ is the vacuum speed of light, $\beta$ is what
we call the propagation constant, $\phi^E(x)$ and $\phi^H(x)$ are the unknown field
distributions. Substituting one of these fields into corresponding Maxwell wave
equation, in the case of permittivity $\epsilon(\bi{r}) = \epsilon(x)$ being
a step-wise function equal to $n_j^2$ for the $j$th homogeneous layer we obtain
the following eigenproblem:
\begin{equation}
\frac{\rmd^2\phi(x)}{\rmd x^2} + \frac{\omega^2}{c^2}(n_j^2 - \beta^2)\phi(x) = 0.
\end{equation}
This is the dispersion equation for a multilayer waveguide; the
difference between TE and TM polarizations lies exclusively in
different boundary conditions at the interfaces of the layers.
Thus, at the interface $x = x_0$ we have:
\begin{eqnarray}
\phi^E(x_0^{-})=\phi^E(x_0^{+}), \qquad &
        \frac{\rmd \phi^E(x_0^{-})}{\rmd x} = \frac{\rmd \phi^E(x_0^{+})}{\rmd x}, \\
\bs     \phi^H(x_0^{-})=\phi^H(x_0^{+}), &
{1\over\epsilon(x_0^{-})}\frac{\rmd \phi^H(x_0^{-})}{\rmd x} = {1\over\epsilon(x_0^{+})}\frac{\rmd \phi^H(x_0^{+})}{\rmd x}.
\end{eqnarray}
Conditions for the derivatives originate from continuity of
$z$-components of magnetic and electric field vectors via Maxwell
equations for $\nabla\times\bi{E}$ and $\nabla\times\bi{H}$
respectively.

Note that the scalar Helmholtz wave equation (7) with boundary
conditions for the TE mode (8) is isomorphous to the familiar
stationary one-dimensional  Schr\"{o}dinger equation
\begin{equation}
\frac{\rmd^2\psi(x)}{\rmd x^2} + \frac{2m}{\hbar^2}({\cal E} - U(x))\psi(x) = 0
\end{equation}
for a particle of mass $m$ in potential energy profile $U(x)$,
$\cal E$ being the energy levels corresponding to eigenfunctions
$\psi_{\cal E}(x)$.

\begin{figure}
\begin{center}
\includegraphics{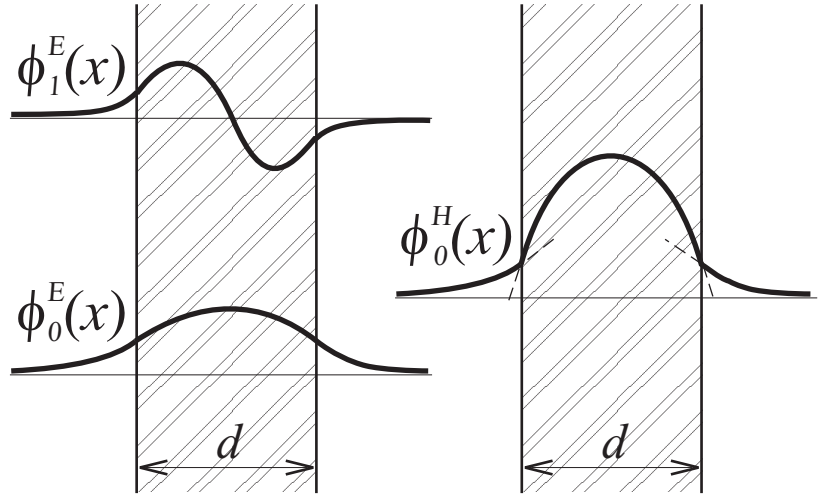}
\end{center}
\caption{Illustration to the rule of alternation of TE and TM
modes in a single-layer waveguide.}
\label{f2}
\end{figure}

The difference in boundary conditions for the derivatives of
$\phi^E(x)$ and $\phi^H(x)$ functions leads to strict alternation
of TE and TM dispersion curves of one- and two-layer waveguides.
In other words, at a fixed frequency $\omega$ the propagation
constants of confined TE modes are always separated by those of TM
ones. This `rule' can be illustrated by
\hyperref[f2]{figure~\ref*{f2}}  for the lowest modes in a
single-layer waveguide: the ratio $\lambda / 2d$ ($\lambda$ is the
wavelength {\em within\/} the guide) for the $\phi^E(x)$  function
of the TE$_0$ mode is obviously more than that for the TM$_0$ mode
because of the steeper slopes of the $\phi^H(x)$ function near the
interfaces due to the boundary conditions, but however high the
refractive index of the guide and consequently the slopes of
$\phi^H(x)$ function in vicinity of the interfaces may be, the
ratio $\lambda / 2d$ for the TM$_0$ mode will never drop below
unity, while for the next TE mode---TE$_1$---it is definitely less
than 1, actually lying between 1/2 and 1.

Similar argumentation still holds for the modal dispersion
structure of a two-layer stack, no matter of the refractive index
contrast of the layers, and, in general, for $N$-layer stack with
refractive indices of the layers regularly increasing or
decreasing throughout the array ($n_1<n_2<\dots<n_N$ or
$n_1>n_2>\dots>n_N$).

Surprisingly, a three-layer waveguide of step-wise refractive
index profile develops considerably different from the above case,
provided the second layer is of lower refractive index then its
surroundings: $n_2 < n_{1,3}$. Also assume for the sake of
simplicity $n_1 = n_3$ and, as before, symmetric sandwiching: $n_0
= 1$ for both ambient media. As it will now be argued, the mode
matching can be effectively implemented with such a system; to
understand its origin, it might be a helpful trick to recall the
changes in the system of eigenfunctions $\psi_{\cal E}(x)$ of an
electron in one-dimensional three-step potential gap, each step
being of width $d/3$:
\begin{equation}
U(x)=\cases{U_0&for $|x| > d/2$ (outside the gap)\\
        U_2&for $|x| < d/6$ (in the middle region)\\
        U_{1,3}&otherwise (in the lateral regions)\\}
\end{equation}
with increasing $U_2$ from $U_2 = U_{1,3}$ to $U_2 = U_0$ that
corresponds to decreasing $n_2$ from $n_2 = n_{1,3}$ (isotropic
single-layer waveguide) to unity (two waveguide channels separated
by air region), the frequency $\omega$ being fixed.

While the difference between $U_{1,3}$ and $U_2$ is negligible,
the energy levels correspondent to $\psi_{\cal E}(x)$
eigenfunctions compose a relatively regular discrete energy
spectrum, and so do the propagation constants of TE modes at any
fixed frequency. The TM mode spectrum is a bit shifted with
respect to the TE one due to the difference in boundary conditions
for the modes of different types, and as just
emphasized, the propagation constants of TE and TM modes are
strictly alternating.

\begin{figure}
\begin{center}
\includegraphics[scale=0.83]{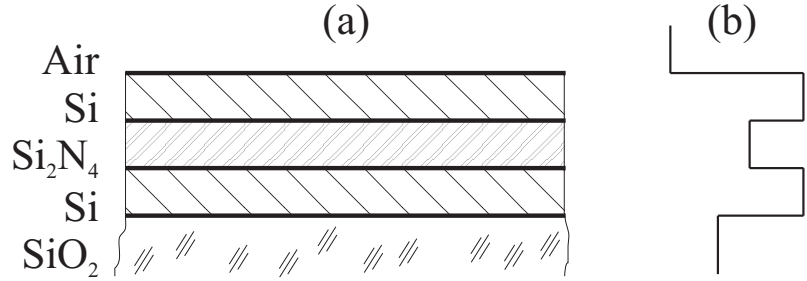}
\end{center}
\caption{(a) Schematic diagram of the three-layer dielectric
waveguide on silica substrate and (b) its refractive index
profile.} \label{f3}
\end{figure}

By contrast, when $U_2$ is large enough, the regions of low
potential become weakly coupled, the entire system possessing the
twofold (generally $N$-fold, where $N$ is the number of regions of
low potential---or high refractive index) splitting of mixed odd
and even energy levels corresponding to nearly degenerate TE modal
{\em groups\/} of $N$ modes each. TM modes also form similar
$N$-fold groups which are, however, shifted with respect to TE
groups. It is clear now that for certain $n_2$ lying between those
extremes, or equivalently, for certain frequency $\omega$ in case
of $n_2$ being somewhat less than $n_{1,3}$ and fixed, the
matching of TE and TM modes should be expected, with the only
exception for TE$_0$ mode in agreement with the inevitably
negative character of form birefringence \cite{BornWolf,GuYeh}. We
can also anticipate herefore that the minimum sufficient number of
layers for the perfect phase matching is three, provided the
central layer is of lower refractive index than the surrounding
ones---see \hyperref[f3]{figure~\ref*{f3}}  for an example of such
a structure. More generally, the refractive index profile of the
waveguide should have more than one maximum for the effective mode
matching.

Consider now a kind of three-layer system `reciprocal' to the
above---with a middle layer being of higher index than the lateral
ones. For this third possible arrangement of the layers, the mode
matching is not forbidden, yet there is seemingly no solid ground
to expect the matching of the lowest modes now since such a system
appears to stand close to conventional slab waveguide---the latter
with no possibility for the mode matching at all. More thorough
analysis in the manner above for a single-layer waveguide shows
that coincidence of propagation constants should be expected for
the modes {\em of the same order\/}, but as before with the
exception of TE$_0$ and TM$_0$ modes. It is demonstrated in
\hyperref[Simulated]{section~\ref*{Simulated}}  that the mode
matching is indeed observed in this case only in the multimode
regime, thus actually losing its practical value.

%%%%%%%%%%%%%%%%%%%%%%%%%%%%%%%%%%%%%%%%%%%%%%%%%%%%%%%%%%%%%

\section{\label{Dispersion}Dispersion analysis for multiperiod two- and three-layer waveguides}
The general form (7)--(9) of the dispersion equations for guided
modes is quite good for qualitative analysis, but inconvenient for
numerical investigations into the TE and TM mode spectra of
multilayer stacks. In the current section we render the known
dispersion relations for multiperiod bilayer waveguides and
present their generalization for three-layer multiperiod systems.
Thus, following Yeh~\cite{Yeh88}, in case of an $N$-period stack
with a {\em bilayer\/} unit cell one has for TE modes:
\begin{eqnarray}
\frac{\sin (N \Lambda_{2s})}{\sin \Lambda_{2s}} \left[ \left( \frac{1}{\xi_{Sub}}+\frac{1}{\xi_{Cov}} \right)
\mathrm{CC} - \left( \frac{\eta_1}{\xi_{Sub}}+\frac{\eta_2}{\xi_{Cov}} \right)
\mathrm{SS} \right. \nonumber\\
\bs  - \left.\left( \frac{\eta_1}{\xi_{Sub}\xi_{Cov}} - {1\over\eta_2} \right)\mathrm{SC}
- \left( \frac{\eta_2}{\xi_{Sub}\xi_{Cov}} - {1\over\eta_1} \right)\mathrm{CS} \right] \nonumber\\
\bs  - \left( \frac{1}{\xi_{Sub}}+\frac{1}{\xi_{Cov}} \right) \frac{\sin [(N-1)\Lambda_{2s}]}{\sin \Lambda_{2s}} = 0,
\end{eqnarray}
where
\begin{equation}
\Lambda_{2s} = \arccos\left[\mathrm{CC}
- \left( {\eta_1\over\eta_2}+{\eta_2\over\eta_1} \right){\mathrm{SS}\over2} \right],
\end{equation}
$\eta_j = \sqrt{n_j^2 - \beta^2}$, $\xi_{Sub,Cov} = \sqrt{\beta^2
- n_{Sub,Cov}^2}$ and the shorthand notation such as $\mathrm{SC}$
stands for the corresponding harmonic function product, in the
case above, $\sin(kt_1n_1)\cos(kt_2n_2)$, with the free-space wave
number $k = \omega/c$.

Likewise for TM modes~\cite{Yeh88}:
\begin{eqnarray}
\frac{\sin (N \Lambda_{2p})}{\sin \Lambda_{2p}} \left[ \left( \frac{\xi_{Sub}}{\epsilon_{Sub}}+\frac{\xi_{Cov}}{\epsilon_{Cov}} \right)
\mathrm{CC} - \left( \frac{\xi_{Sub}\epsilon_1\eta_2}{\epsilon_{Sub}\epsilon_2\eta_1}+\frac{\xi_{Cov}\epsilon_2\eta_1}{\epsilon_{Cov}\epsilon_1\eta_2}
\right)\mathrm{SS} \right. \nonumber\\
\bs  \left. - \left( \frac{\eta_1}{\epsilon_1} + \frac{\xi_{Sub}\xi_{Cov}\epsilon_1}{\epsilon_{Sub}\epsilon_{Cov}\eta_1}\right)\mathrm{SC}
- \left( \frac{\eta_2}{\epsilon_2} + \frac{\xi_{Sub}\xi_{Cov}\epsilon_2}{\epsilon_{Sub}\epsilon_{Cov}\eta_2} \right)\mathrm{CS} \right]\nonumber\\
\bs  - \left( \frac{\xi_{Sub}}{\epsilon_{Sub}}+\frac{\xi_{Cov}}{\epsilon_{Cov}} \right) \frac{\sin [(N-1)\Lambda_{2p}]}{\sin \Lambda_{2p}} = 0,
\end{eqnarray}
where
\begin{equation}
\Lambda_{2p}= \arccos\left[\mathrm{CC} - \left( \frac{\eta_1\epsilon_2}{\eta_2\epsilon_1}+\frac{\eta_2\epsilon_1}{\eta_1\epsilon_2} \right){\mathrm{SS}\over2} \right].
\end{equation}

The above equations look rather awkward, but they can be easily
handled and promptly solved numerically in any mathematical
computing package.

Three-layer semi-infinite dielectric superlattices have been
studied earlier, but with the focus on the existence and the
dispersion of bulk and surface polaritons~\cite{Mendialdua}. In
the case of a {\em finite\/}, $N$-period stack with a trilayer
unit cell one can however quite trivially derive the following
dispersion relation for the guided TE modes (see \hyperref
[append]{the appendix} for the details):
\begin{eqnarray}
\frac{\sin (N \Lambda_{3s})}{\sin \Lambda_{3s}} \left[ \left( \frac{1}{\xi_{Sub}}+\frac{1}{\xi_{Cov}}\right)\mathrm{CCC}
- \left( \frac{\eta_2}{\eta_1\eta_3}+{1\over{\xi_{Sub}\xi_{Cov}}}\frac{\eta_1\eta_3}{\eta_2}
\right)\mathrm{SSS} \right. \nonumber\\
\bs  - \left( {1\over\xi_{Sub}}{\eta_1\over\eta_2} + {1\over\xi_{Cov}}{\eta_2\over\eta_1}
\right)\mathrm{SSC} - \left({1\over\xi_{Sub}}{\eta_1\over\eta_3} + {1\over\xi_{Cov}}{\eta_3\over\eta_1}\right)\mathrm{SCS} \nonumber\\
\bs  - \left({1\over\xi_{Sub}}{\eta_2\over\eta_3} + {1\over\xi_{Cov}}{\eta_3\over\eta_2}\right)\mathrm{CSS}
- \left( {\eta_1\over\xi_{Sub}\xi_{Cov}}+{1\over\eta_1} \right)\mathrm{SCC} \nonumber\\
\bs  - \left.\left( {\eta_2\over\xi_{Sub}\xi_{Cov}}+{1\over\eta_2} \right)\mathrm{CSC}
- \left( {\eta_3\over\xi_{Sub}\xi_{Cov}}+{1\over\eta_3} \right)\mathrm{CCS} \right] \nonumber\\
- \left( \frac{1}{\xi_{Sub}}+\frac{1}{\xi_{Cov}} \right) \frac{\sin [(N-1)\Lambda_{3s}]}{\sin \Lambda_{3s}} = 0,
\end{eqnarray}
where
\begin{eqnarray}
\Lambda_{3s}= \arccos\left[\mathrm{CCC} - \left( {\eta_1\over\eta_2}+{\eta_2\over\eta_1}\right){\mathrm{SSC}\over2}\right. \nonumber\\
\left. - {1\over2}\left( {\eta_1\over\eta_3}+{\eta_3\over\eta_1} \right){\mathrm{SCS}\over2}
- {1\over2}\left( {\eta_2\over\eta_3}+{\eta_3\over\eta_2} \right){\mathrm{CSS}\over2} \right],
\end{eqnarray}
and for TM modes:
\begin{eqnarray}
\frac{\sin (N \Lambda_{3p})}{\sin \Lambda_{3p}} \left[ \left( \frac{\xi_{Sub}}{\epsilon_{Sub}}+\frac{\xi_{Cov}}{\epsilon_{Cov}} \right) \mathrm{CCC}
-\left(\frac{\xi_{Sub}\xi_{Cov}}{\epsilon_{Sub}\epsilon_{Cov}} \frac{\epsilon_1\eta_3}{\epsilon_2} \frac{\eta_2}{\eta_1\eta_3}
+\frac{\epsilon_2}{\epsilon_1\eta_3} \frac{\eta_1\eta_3}{\eta_2}
\right)\mathrm{SSS} \right. \nonumber\\
\bs  - \left( \frac{\xi_{Sub}\epsilon_1\eta_2}{\epsilon_{Sub}\epsilon_2\eta_1}+\frac{\xi_{Cov}\epsilon_2\eta_1}{\epsilon_{Cov}\epsilon_1\eta_2} \right)
\mathrm{SSC} - \left( \frac{\xi_{Sub}\epsilon_1\eta_3}{\epsilon_{Sub}\epsilon_3\eta_1}+\frac{\xi_{Cov}\epsilon_3\eta_1}{\epsilon_{Cov}\epsilon_1\eta_3} \right)
\mathrm{SCS} \nonumber\\
\bs  - \left( \frac{\xi_{Sub}\epsilon_2\eta_3}{\epsilon_{Sub}\epsilon_3\eta_2}+\frac{\xi_{Cov}\epsilon_3\eta_2}{\epsilon_{Cov}\epsilon_2\eta_3} \right)
\mathrm{CSS} - \left( \frac{\eta_1}{\epsilon_1} + \frac{\xi_{Sub}\xi_{Cov}\epsilon_1}{\epsilon_{Sub}\epsilon_{Cov}\eta_1} \right)
\mathrm{SCC} \nonumber\\
\bs  \left. - \left( \frac{\eta_2}{\epsilon_2} + \frac{\xi_{Sub}\xi_{Cov}\epsilon_2}{\epsilon_{Sub}\epsilon_{Cov}\eta_2} \right)
\mathrm{CSC} - \left( \frac{\eta_3}{\epsilon_3} + \frac{\xi_{Sub}\xi_{Cov}\epsilon_3}{\epsilon_{Sub}\epsilon_{Cov}\eta_3} \right)
\mathrm{CCS} \right] \nonumber\\
\bs  - \left( \frac{\xi_{Sub}}{\epsilon_{Sub}}+\frac{\xi_{Cov}}{\epsilon_{Cov}} \right) \frac{\sin [(N-1)\Lambda_{3p}]}{\sin \Lambda_{3p}} = 0,
\end{eqnarray}
where
\begin{eqnarray}
\Lambda_{3p}= \arccos\left[\mathrm{CCC} - \left( \frac{\eta_1\epsilon_2}{\eta_2\epsilon_1}+\frac{\eta_2\epsilon_1}{\eta_1\epsilon_2} \right)
{\mathrm{SSC}\over2} \right. \nonumber\\
- \left.\left( \frac{\eta_1\epsilon_3}{\eta_3\epsilon_1}+\frac{\eta_3\epsilon_1}{\eta_1\epsilon_3} \right)
{\mathrm{SCS}\over2} - \left( \frac{\eta_2\epsilon_3}{\eta_3\epsilon_2}+\frac{\eta_3\epsilon_2}{\eta_2\epsilon_3} \right)
{\mathrm{CSS}\over2} \right].
\end{eqnarray}

Of course, certain symmetries could obviously be found in these
equations, but still the latter seem to be too cumbersome for any
qualitative speculations. Instead, one can easily do now plenty of
computational work related to physically realizable multilayers,
and in the following section we will do that with waveguides made
of such popular optical materials as silicon and its compounds.

%%%%%%%%%%%%%%%%%%%%%%%%%%%%%%%%%%%%%%%%%%%%%%%%%%%%%%%%%%%%%

\section{\label{Simulated}Simulated results for multilayers of silicon and silicon nitride}
\begin{figure}
\begin{tabular}{cc}
\includegraphics[scale=0.65]{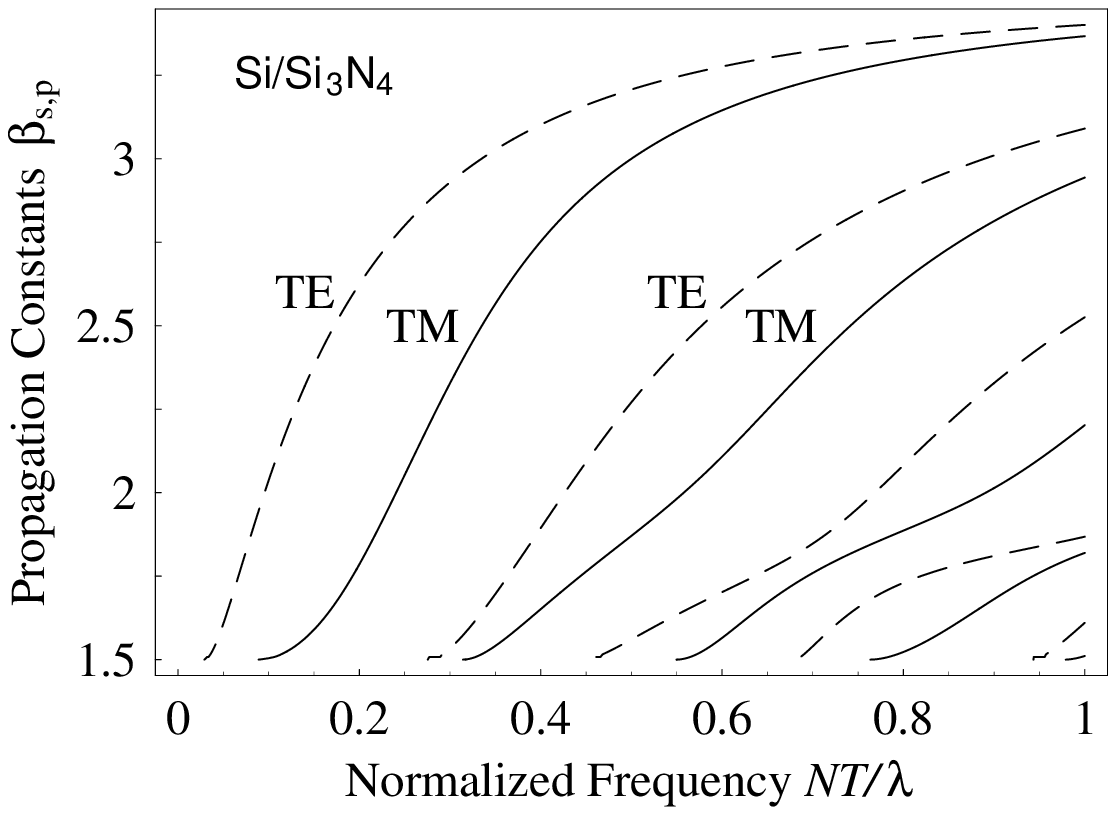}&\includegraphics[scale=0.65]{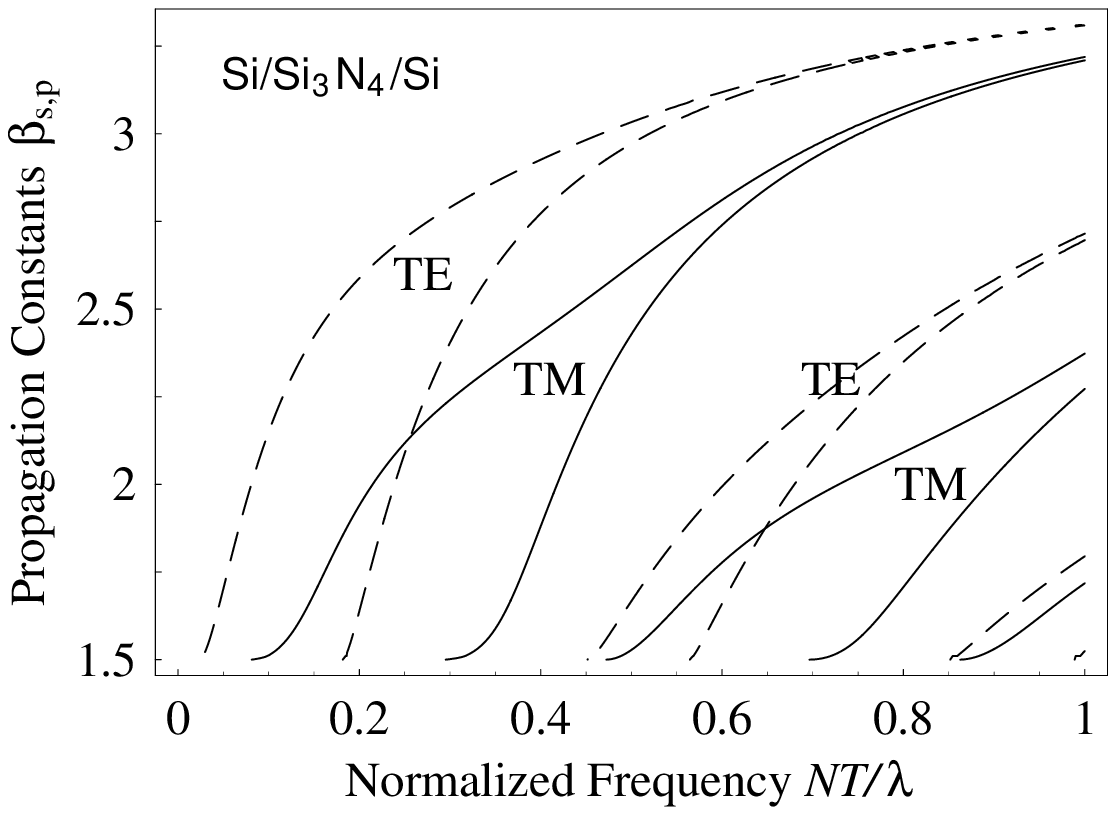}\\
\includegraphics[scale=0.65]{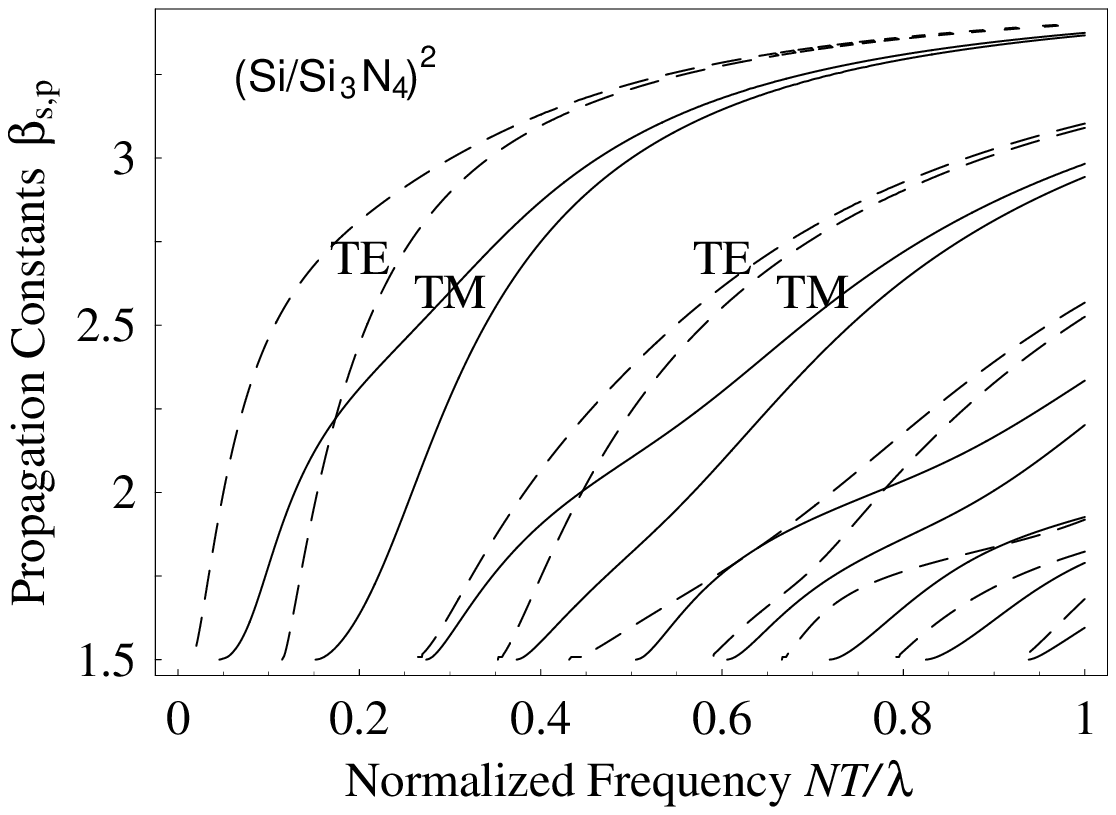}&\includegraphics[scale=0.65]{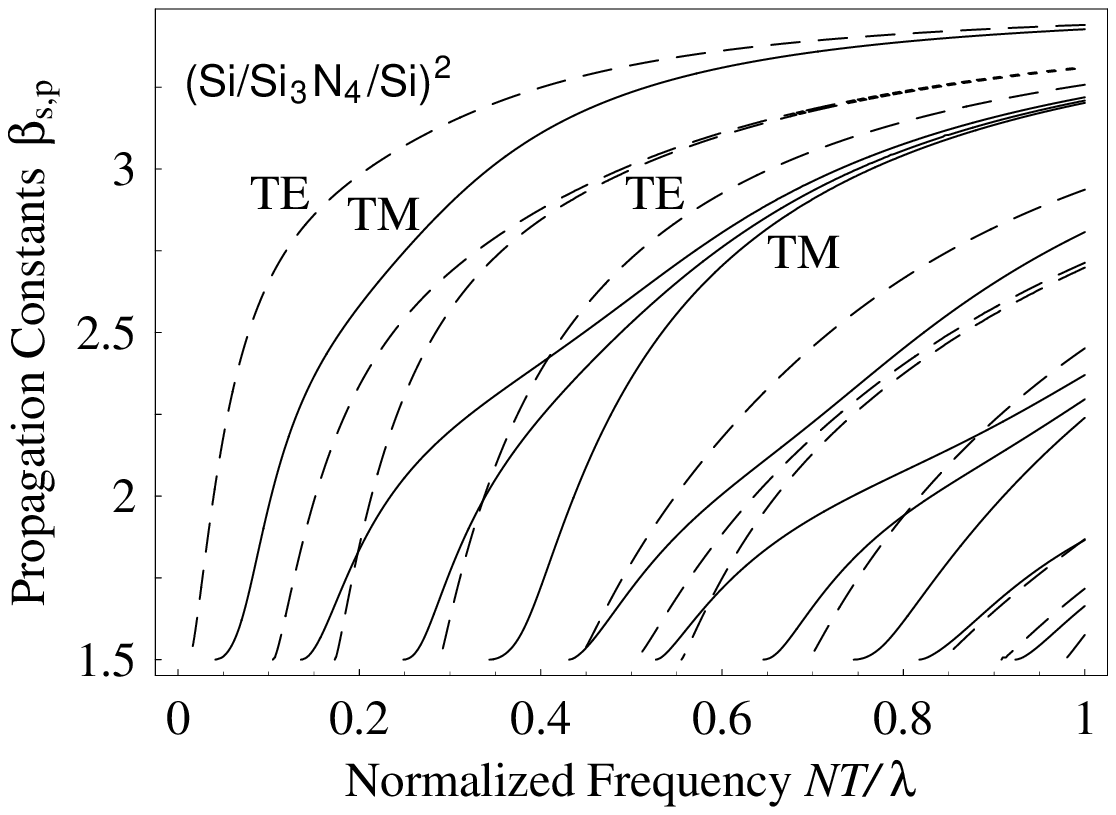}
\end{tabular}
\caption{TE (\broken) and TM (\full) families of the modal
dispersion curves for (a) bilayer ${\rm Si/Si_3N_4}$, (b)
three-layer ${\rm Si/Si_3N_4/Si}$, (c) two-period four-layer ${\rm
(Si/Si_3N_4)^2}$ and (d) five-layer ${\rm (Si/Si_3N_4/Si)^2}$
waveguides, all on silica substrate.} \label{f4}
\end{figure}

To demonstrate the performance of dielectric multilayers for the
mode matching and to corroborate theoretical considerations
presented in \hyperref[WhatIs]{section~\ref*{WhatIs}}, we have
plotted in \hyperref[f4]{figure~\ref*{f4}} the modal dispersion
curves of (a) two-layer ${\rm Si/Si_3N_4}$ and (b) three-layer
${\rm Si/Si_3N_4/Si}$ waveguides, as well as of (c,d) their
two-period counterparts. Silicon and its nitride have been chosen
deliberately since these materials are found to be suitable for
manifold passive optical components like branching waveguides,
couples, interconnects, interferometers, power splitters, and
filters \cite{Worhoff,Schauwecker}. We employed here the mode
dispersion relations for bilayer and three-layer waveguides and
their generalization for multiperiod systems, equations
(12)--(19). The thicknesses of the layers were assumed to be
equal; the refractive indices are 3.5 (Si), 2.0 (Si$_3$N$_4$) and
1.5 (SiO$_2$). All the numerical calculations were performed with
Mathematica 4.1~\cite{Mathematica}.

Figure 4(a) is of no particular interest---we see the system of TE
and TM dispersion curves that is quite typical for slab dielectric
waveguides, with just slight perturbations of the high-order modal
curves near $\beta\approx2$---that is close to the refractive
index of the middle layer. In contrast, figure 4(b) reveals the
predicted intersection of modal dispersion curves for the
three-layer system; for the first phase-matching point where the
TM$_0$ mode matches TE$_1$ at $NT=0.25\lambda$ (here $T$ is the
total thickness of one period, so that $NT$ gives the thickness of
the whole structure), we can estimate the thicknesses of the
layers to be $t_1=t_2=t_3=130$~nm for operation near 1.55~$\mu$m,
i.e. in the spectral range for telecommunication frequencies. In
order to improve the operation bandwidth, the slopes of dispersion
curves at the given point of intersection must be as close as
possible, and for that purpose the refractive index distribution
should be further optimized.

\begin{figure}
\begin{center}
\includegraphics[scale=0.65]{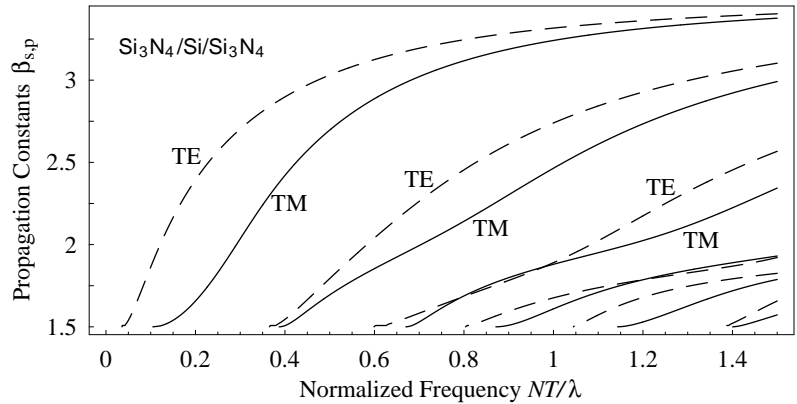}
\end{center}
\caption{Modal dispersion curves for three-layer ${\rm
Si_3N_4/Si/Si_3N_4}$ waveguide on silica substrate---the `reverse'
structure of that in \hyperref[f4]{figure~4(b)}.} \label{f5}
\end{figure}

The comparison between figures 4(b), (c) and (d) confirms that it
is the number of layers of high refractive index, and not
periods---which sometimes are more `mental' constructs than
`physical' objects---that determines the number of modes in the TE
and TM bands formed. Thus, for both ${\rm Si/Si_3N_4/Si}$ and
${\rm (Si/Si_3N_4)^2}$ structures constituted of two silicon
layers---two `potential gaps' for photons---the twofold grouping
of modes at high $\beta$ values is clearly observed. Finally,
\hyperref[f5]{figure~\ref*{f5}} testifies to the statement of
\hyperref[WhatIs]{section~\ref*{WhatIs}} that for the effective
mode matching, the refractive index of the medium layer in
three-layer waveguide should be low enough; otherwise, the
matching is not forbidden, but occurs only for high-order modes
and remains thus impracticable.

%%%%%%%%%%%%%%%%%%%%%%%%%%%%%%%%%%%%%%%%%%%%%%%%%%%%%%%%%%%%%

\section{Conclusion}
In summary, the perfect phase matching of TE and TM guided modes
with a multilayer waveguide composed of linear isotropic
dielectric materials has been anticipated and simulated for the
first time. Dielectric superlattices with unit cells of two types
have been considered: two-layer and tree-layer; for finite,
$N$-period, three-component structures, the compact analytical
expressions for the mode dispersion have been derived, which can
be regarded as a self-sufficient result.

Numerical simulations performed for a
silicon---silicon-nitride---silicon waveguide on a silica
substrate bear witness that even a three-layer stack can be
designed to exhibit at certain frequencies the perfect phase
matching of TE and TM guided modes. Since the TE and TM mode
spectra are naturally very sensitive to refractive index
distribution, i.e.~to the index contrast of the layers and their
thicknesses, number and succession, wide possibilities arise to
tailor the mode dispersion in the vicinity of those phase-matching
points.

The principle of perfect phase matching with a multilayer
dielectric array has important potential implementations in
polarization-insensitive planar propagation, fibre-to-planar
waveguide coupling, mode conversion, filtering etc. The advantage
of the presented device scheme is its simplicity and compatibility
with common integrated optics techniques, e.g. with planar optical
waveguide technology. Further important issues to be addressed for
the development of applicational aspects of the scheme are
optimization with a more advanced refractive index profile of the
waveguiding stack and integration into complex, state-of-the-art
optical and optoelectronic elements.

%%%%%%%%%%%%%%%%%%%%%%%%%%%%%%%%%%%%%%%%%%%%%%%%%%%%%%%%%%%%%

\appendix
\section*{\label{append}Appendix}
\setcounter{section}{1} To derive the mode dispersion relation for
the finite, $N$-period, three-layer dielectric superlattices, one
can simply follow Yeh's~\cite{Yeh88} reasoning for bilayer
multiperiod waveguides, yet we prefer to utilize another tool
here---the general, dyadic-based formalism for stratified
media~\cite{BBL87} which proves to be a competitive modification
of Berreman $4\times4$ matrix approach~\cite{Berreman72} and can
be directly applied to more complicated problems of light
scattering and confined propagation in, e.g., magnetized,
metallo-dielectric and anisotropic arrays.

Consider an $N$-period three-layer stack sandwiched between
isotropic, homogeneous, semi-infinite media---a substrate of
permittivity $\epsilon_{Sub}$ and a cover of permittivity
$\epsilon_{Cov}$. The covariant dispersion relation for
plane-stratified media~\cite{BBL87} takes the form
\begin{equation}
\det \left[ (-\gamma_{Sub},\;{\cal I})\times({\cal P}_1{\cal P}_2{\cal P}_3)^N
\times\left( \begin{array}{c}{\cal I}\\-\gamma_{Cov}\end{array} \right)\right]=0,
\end{equation}
where ${\cal I}=\bi z\otimes\bi z+\bi y\otimes\bi y$ is the unit
2-D dyadic, $\bi z$ and $\bi y$ are the unit vectors in the plane
of the guide, respectively parallel and perpendicular to the
direction of propagation, $\bi a\otimes\bi b$ denotes the outer
product $a_\alpha b_\beta$, ${\cal P}_j$ is the evolution operator
($4\times4$ matrix) that relates $z$ and $y$ components of
electric and magnetic fields arranged into truncated $\bi
x\times\bi E=(-E_z,E_y)^T$ and $\bi H_\tau=(H_y,H_z)^T$ vectors at
the interfaces $x=x_j$ and $x=x_{j-1}$ of the $j$th layer,
\begin{equation}
\left( \begin{array}{c}\bi{H}_\tau(x_j)\\ \bi x\times\bi E(x_j)\end{array} \right)
={\cal P}_j \left( \begin{array}{c}\bi{H}_\tau(x_{j-1})\\ \bi x\times\bi E(x_{j-1})\end{array}\right),
\end{equation}
and $\gamma_{Sub}$, $\gamma_{Cov}$ are the impedance tensors of
the ambient media that allow to generate $\bi H_\tau$ given $\bi
x\times\bi E$: $\bi H_\tau=\gamma\,(\bi x\times\bi E)$,
\begin{equation}
\gamma_{Sub,Cov} = {1\over{\rmi\xi_{Sub,Cov}}}\,\bi{z}\otimes\bi{z}
+\frac{\rmi\xi_{Sub,Cov}}{\epsilon_{Sub,Cov}}\,\bi{y}\otimes\bi{y}.
\end{equation}

In the case of an isotropic dielectric layer, the evolution matrix
is
\begin{eqnarray}
{\cal P}_j = {^z\cal P}_j \bi{z}\otimes\bi{z} + {^y\cal P}_j\bi{y}\otimes\bi{y}\nonumber\\
\lo=\left( \begin{array}{cc}\mathrm C_j&\rmi\eta_j\mathrm S_j\\(\rmi/\eta_j)\mathrm S_j&\mathrm C_j\end{array} \right)
\bi{z}\otimes\bi{z}
+\left( \begin{array}{cc}\mathrm C_j&(\rmi\epsilon_j/\eta_j)\mathrm S_j\\(\rmi\eta_j/\epsilon_j)\mathrm S_j&\mathrm C_j\end{array} \right)
\bi{y}\otimes\bi{y}.
\end{eqnarray}

Such a simple structure of operators ${\cal P}_j$ and
$\gamma_{Sub,Cov}$ permits us to decompose dispersion equation
(A1) into two separate equations---for TE and TM modes; these are
obtained via multiplying $^z{\cal P}_1$, $^z{\cal P}_2$, $^z{\cal
P}_3$ or $^x{\cal P}_1$, $^x{\cal P}_2$, $^x{\cal P}_3$ and
applying the known representation of $N$th power of a $2\times2$
unimodular matrix in terms of the Chebychev polinomials of the
second kind given by Abel\`{e}s~\cite{Abeles}. Finally, this
cumbersome, but physically transparent procedure results in
equations (16)--(19).

%%%%%%%%%%%%%%%%%%%%%%%%%%%%%%%%%%%%%%%%%%%%%%%%%%%%%%%%%%%%%


\section*{References}
\begin{thebibliography}{99}

\bibitem{Tien74}Tien P K, Schinke D P and Blank S L 1974 Magneto-optics and motion of the magnetization in a film-waveguide optical switch
        \textit{J. Appl. Phys.} \textbf{45} 3059--68

\bibitem{Bruns75}Bruns W K and Milton A F 1975 Mode conversion in planar-dielectric separating waveguides
        \textit{IEEE J. Quantum Electron.} \textbf{11} 32--9

\bibitem{Gillies76}Gillies J R and Hlawiczka P 1976 TE and TM modes in gyrotropic waveguides
        \textit{\JPD}\textbf{9} 1315--22

\bibitem{Hlawiczka78a}Hlawiczka P 1978 A gyrotropic waveguide with dielectric boundaries: the longitudinally magnetised case
        \textit{\JPD}\textbf{11} 1157--66

\bibitem{Hlawiczka78b}Hlawiczka P 1978 The gyrotropic waveguide with a normal applied DC field
        \textit{\JPD}\textbf{11} 1941--8

\bibitem{Winn}Winn J N, Fink Y, Fan S, Joannopoulos J D 1998 Omnidirectional reflection from a one-dimensional photonic crystal
        \textit{Opt. Lett.} \textbf{23} 1573--5

\bibitem{Dowling}Dowling J P 1998 Mirror on the wall: you're omnidirectional after all?
        \textit{Science} \textbf{288} 1841--2

\bibitem{Chigrin}Chigrin D N, Lavrinenko A V, Yarotsky D A and Gaponenko S V 1999 Observation of total omnidirectional reflection from a one-dimensional dielectric lattice
        \textit{Appl. Phys.} A \textbf{68} 25--8

\bibitem{Russell}Russell P St J, Tredwell S and Roberts R J 1999 Full photonic band-gaps and spontaneous emission control in 1-D multilayer dielectric structures
        \textit{Opt. Commun.} \textbf{160} 66--71

\bibitem{Joannopoulos}Joannopoulos J D, Meade R D and Winn J N 1995 \textit{Photonic Crystals: Molding the Flow of Light}
        (Princeton, NJ: Princeton University Press)

\bibitem{me}Lavrinenko A V, Shyroki D N and Zhilko V V 2002 Guided and leaky modes in magneto-optical sandwiches with ultrathin metal films
        \textit{J. Magn. Magn. Mater.} \textbf{247} 171--7

\bibitem{BornWolf}Born M and Wolf E 1968 4th edn \textit{Principles of Optics}
        (Oxford: Pergamon) ch~14

\bibitem{GuYeh}Gu C and Yeh P 1996 Form birefringence dispersion in periodic layered media
        \textit{Opt. Lett.} \textbf{21} 504--6

\bibitem{Yeh88} Yeh P 1988 \textit{Optical Waves in Layered Media}
        (New York: Wiley) ch~11

\bibitem{Mendialdua}Mendialdua J, Rodriguez A, More M, Akjouj A and Dobrzynski L 1994 Bulk and surface phonon polaritons in three-layer superlattices
        \textit{Phys. Rev.} B \textbf{50} 14 605--8

\bibitem{Worhoff}W\"{o}rhoff K, Lambeck P V and Driessen A 1999 Design, tolerance analysis, and fabrication of silicon oxynitride based planar waveguides for communication devices
        \textit{J. Lightwave Technol.} \textbf{17} 1401--7

\bibitem{Schauwecker}Schauwecker B, Arnold M, Radehaus C V, Przyrembel G and Kuhlow B 2002 Optical waveguide components with high refractive index difference in silicon-oxynitride for application in integrated optoelectronics
        \textit{Opt. Eng.} \textbf{41} 237--43

\bibitem{Mathematica}Wolfram research (\url{www.wolfram.com})

\bibitem{BBL87}Barkovskii L M, Borzdov G N and Lavrinenko A V 1987 Fresnel's reflection and transmission operators for stratified gyroanisotropic media
        \textit{J. Phys. A: Math. Gen.} \textbf{20} 1095--106

\bibitem{Berreman72}Berreman D W 1972 Optics in stratified and anisotropic media: $4\times4$-matrix formulation
        \textit{J. Opt. Soc. Am} \textbf{62} 502--10

\bibitem{Abeles}Abel\`{e}s F 1950 Reserches sur la propagation des ondes electromagnetiques sinusoidales dans les milieux stratifi\`{e}s. Application aux couches minces
        \textit{Ann. Physique} \textbf{5} 596--640

\end{thebibliography}
\end{document}